\documentclass[twocolumn,nofootinbib]{aastex7} %,linenumbers
\usepackage{natbib}
\usepackage{bm}% bold math
\usepackage{color}
\usepackage{graphicx}                           
\usepackage{float}
\usepackage{amsmath}	% Advanced maths commands
\usepackage{multirow} 
\usepackage{amssymb}	% Extra maths symbols
\usepackage{diagbox}

\begin{document}

\title{A New Wavelet Scattering Transform-Based Statistic for Cosmological Analysis of Large-Scale Structure}%

\author{Zhujun Jiang}
\affiliation{School of Physics and Astronomy, Sun Yat-Sen University, Zhuhai 519082, China}
\email{jiangzhj26@mail2.sysu.edu.cn}

\author{Xiaolin Luo}
\affiliation{Department of Astronomy, Shanghai Jiao Tong University, 800 Dongchuan Road, Shanghai 200240, China}
\email{}

\author{Wenying Du}
\affiliation{School of Physics and Astronomy, Sun Yat-Sen University, Zhuhai 519082, China}
\email{}

\author{Zhiwei Min}
\affiliation{School of Physics and Astronomy, Sun Yat-Sen University, Zhuhai 519082, China}
\email{}

\author{Fenfen Yin}
\affiliation{School of Physics and Astronomy, Sun Yat-Sen University, Zhuhai 519082, China}
\email{}

\author{Longlong Feng}
\affiliation{School of Physics and Astronomy, Sun Yat-Sen University, Zhuhai 519082, China}
\email{}

\author{Jiacheng Ding}
\affiliation{Shanghai Astronomical Observatory (SHAO), Nandan Road 80, Shanghai 200030, China}
\email[show]{dingjch@shao.ac.cn}

\author{Le Zhang}
\email[show]{zhangle7@mail.sysu.edu.cn}
\affiliation{School of Physics and Astronomy, Sun Yat-Sen University, Zhuhai 519082, China}
\affiliation{CSST Science Center for the Guangdong–Hong Kong–Macau Greater Bay Area, SYSU, Zhuhai 519082, China}

\author{Xiao-Dong Li}
\email[show]{lixiaod25@mail.sysu.edu.cn}
\affiliation{School of Physics and Astronomy, Sun Yat-Sen University, Zhuhai 519082, China}
\affiliation{Peng Cheng Laboratory, Shenzhen, Guangdong 518066, China}
\affiliation{CSST Science Center for the Guangdong–Hong Kong–Macau Greater Bay Area, SYSU, Zhuhai 519082, China}

\begin{abstract}
Large-scale structure (LSS) analysis in galaxy surveys is a powerful cosmological probe but is limited by tracer bias, which can obscure underlying information and weaken parameter constraints. Existing methods either model bias or restrict analyses to low-density regions, yet their sensitivity to bias remains poorly understood. We propose a novel method based on the wavelet scattering transform (WST) to distinguish LSS across cosmological models while mitigating tracer bias. Central to our approach are the WST $m$-mode ratios, $R^{\rm wst}$, a new statistical measure, and a high-density apodization preprocessing that smoothly rescales extreme values. We use a reduced chi-square to assess the cosmological parameter constraints and find that $R^{\rm wst}$, in the scale range $j \in [3,7]$, achieves $\chi^2_{\nu, \rm cos} \approx 6$ for cosmology while maintaining $\chi^2_{\nu, \rm bias} \sim 1$--a regime unattained by other statistics. $R^{\rm wst}$ thus provides robust cosmological sensitivity with effective bias mitigation for future surveys.

\end{abstract}

\section{Introduction}\label{sec:intro}
Understanding the nature of dark energy (DE) and dark matter (DM) remains one of the most important challenges in modern cosmology. The large-scale structure (LSS) of galaxy clusters encodes rich information about the expansion history and the growth of cosmic structures, serving as a crucial probe of the physical nature of DE and DM~\citep{2013PhR...530...87W}.

Over the past two decades, Stage-III galaxy surveys, including 2dFGRS, 6dF, WiggleZ, and SDSS~\citep{2df:Colless:2003wz, beutler_6df_2012, blake2011wigglez, blake2011wigglezb, york2000sloan, Eisenstein:2005su, Percival:2007yw, anderson2012clustering, alam2017clustering, beutler2016clustering}, have mapped millions of galaxies over volumes of several Gpc$^3$, enabling percent-level measurements of baryon acoustic oscillations, precise constraints on the growth rate, and refined estimates of cosmological parameters. Looking ahead, the next generation of Stage-IV surveys, such as those by the Dark Energy Spectroscopic Instrument (DESI)\footnote{https://desi.lbl.gov/}~\citep{levi_desi_2013}, the Vera C. Rubin Observatory (LSST)\footnote{https://www.lsst.org/}~\citep{collaboration_lsst_2009}, the Euclid satellite\footnote{http://sci.esa.int/euclid/}~\citep{laureijs_euclid_2011}, the Roman Space Telescope\footnote{https://roman.gsfc.nasa.gov/}~\citep{eifler_cosmology_2021}, and the Chinese Space Station Telescope (CSST)\footnote{http://nao.cas.cn/csst/}~\citep{zhan_consideration_2011}, will extend this progress even further.

The LSS of the Universe contains rich cosmological information, including the baryon acoustic oscillation (BAO) signal~\citep{eisenstein_detection_2005, gaztanaga_first_2009, percival_baryon_2010, eisenstein_sdss-iii_2011, dawson_sdss-iv_2016, Nunes:2020hzy, nunes_cosmological_2020, philcox_combining_2020, calderon_desi_2024} and redshift-space distortions (RSD)~\citep{kaiser_clustering_nodate, guzzo_test_2008, song_reconstructing_2009, alam_clustering_2017, nunes_arbitrating_2021, de_mattia_completed_2020, bautista_completed_2020, neveux_completed_2020, yu_rsd_2023}, making it a key probe for constraining cosmological parameters.

LSS surveys typically use galaxies as tracers\footnote{{Other tracers such as quasars, Lyman-$\alpha$ forests, 21 cm emission, gravitational lensing and CMB lensing, see e.g.,}~\citep{weinberg_lyman-alpha_2003, lewis_weak_2006, erben_cfhtlens_2013, xu_forecasts_2014, ata_clustering_2018, heymans_kids-1000_2021, collaboration_detection_2023, paul_first_2023, pan_measurement_2023}}, which reside in dark matter halos. However, halos are themselves biased tracers of the matter distribution~\citep{2010ApJ...724..878T}, and galaxy formation efficiency varies with halo mass~\citep{Benson:1999mva}, introducing additional bias. As a result, the galaxy distribution reflects a scale-dependent bias influenced by both halo clustering and baryonic physics, which can be modeled statistically on large scales in the context of perturbation theory~\citep{Desjacques:2016bnm}.

Accurate modeling of tracer bias is essential for interpreting current and upcoming LSS surveys, as tracer bias introduces systematic errors in cosmological parameter estimation. Mitigating this bias is key to achieving robust and precise constraints, and various methodologies have been developed to address it. One common approach uses bias models--such as linear, nonlinear, conditional luminosity function, and halo occupation distribution models--to describe the galaxy–matter relationship~\citep{berlind_halo_2003, van_den_bosch_towards_2007, cacciato_galaxy_2009, van_den_bosch_cosmological_2013, zentner_galaxy_2014,kitaura_clustering_2016, hearin_introducing_2016, desjacques_large-scale_2018, yuan_desi_2024}, though these often rely on simplifying assumptions. Another method treats bias as a free parameter, jointly fitted with cosmological parameters~\citep{cabass_constraints_2022, springel_first_2018, philcox_boss_2022}. Alternatively, bias-independent techniques avoid relying on galaxy clustering altogether, using observables like weak gravitational lensing~\citep{chae_constraints_2002, chae_constraints_2004, giblin_kids-1000_2021}, the cosmic microwave background~\citep{kneller_how_2001, gupta_parameter_2010, galli_cmb_2014, malinovsky_cosmological_2008}, and cross-correlations between CMB lensing and galaxy distributions~\citep{abbott_dark_2022, krolewski_unwise_2020, white_cosmological_2022}. Despite these efforts, fully removing the impact of tracer bias remains a major challenge in precision cosmology.

Extracting non-Gaussian information from the non-linear regime of structure formation is key to fully exploiting the rich data from upcoming LSS surveys. Traditional two-point statistics are highly effective at capturing Gaussian signals but are blind to phase correlations and higher-order structures. To address this limitation, early work based on the discrete wavelet transform (DWT) demonstrated the potential of multiscale decompositions for probing non-Gaussian features in the matter distribution~\citep{Fang:2000by,Feng:2000ih,romeo_n-body_2003,romeo_wavelet_2004,romeo_discreteness_2008}. More recently, the wavelet scattering transform (WST)~\citep{mallat_group_2012} has provided a systematic framework that extends these ideas: by combining multiscale localization with higher-order moments, the WST enables an efficient characterization of non-linear and non-Gaussian structures, thereby going beyond the descriptive power of two-point methods.

WST has been applied across a range of astrophysical contexts, including statistical characterization of non-Gaussian structures in the interstellar medium~\citep{heymans_kids-1000_2021}, probing the cold neutral medium through HI emission morphology~\citep{lei_probing_2023} and analyzing dust polarization maps~\citep{regaldo-saint_blancard_statistical_2020}. It has also been used as an interpretable low-dimensional statistic for 2D non-Gaussian fields, yielding tighter constraints on several cosmological parameters~\citep{allys_new_2020}. Recent applications include 21 cm lightcone generation during the epoch of reionization~\citep{hothi_generative_2025}, studies of the 21 cm forest~\citep{shimabukuro_analyzing_2025}, and detection of the cosmic 21 cm signal~\citep{greig_detecting_2023, greig_exploring_2022}. In three dimensions, WST has been employed to infer cosmological parameters from weak lensing~\citep{cheng_new_2020}, and enhance the extraction of information from stochastic fields~\citep{cheng_how_2021}. Further developments include applications to actual galaxy observations, such as a WST analysis of the BOSS CMASS dataset~\citep{valogiannis_going_2022,blancard_galaxy_2024}, the first use on simulated 3D density fields for cosmological parameter inference~\citep{valogiannis_towards_2022}, exploration of 3D WST coefficients for line-intensity mapping~\citep{chung_exploration_2022}, and constraints on primordial non-Gaussianity~\citep{peron_constraining_2024}.

However, a major challenge in LSS analysis applying WST is the uncertainty introduced by tracer bias, which can obscure cosmological information and limit the robustness of parameter constraints. While some studies have attempted to model bias or restrict analyses to low-density regions, the sensitivity of current statistical approaches to tracer bias is not yet fully understood~\citep{valogiannis_going_2022, valogiannis_towards_2022,Valogiannis:2023mxf}. To address this gap, we introduce a new statistic -- the $m$-mode ratios of WST coefficients -- that reduces the impact of tracer bias without requiring explicit bias modeling. This method preserves the ability to distinguish LSS realizations across different cosmological parameters while mitigating bias-induced systematics.

The paper is organized as follows. We present WST and the WST $m$-mode ratios we proposed in Sect.~\ref{sec:Scattering Transforms}. We present simulation datasets based on two cosmological models used to validate our method and high-density apodization in Sect.~\ref{sec:Simulation datasets}. The chi-square statistic is presented in Sect.~\ref{sec:chi-square statistic}. The analysis and results are presented in Sect.~\ref{sec:results}. Finally, we summarizes the findings and outlines future directions in Sect.~\ref{sec:conclusion}.

\section{Wavelet Scattering Transforms}\label{sec:Scattering Transforms} 
We first introduce WST in Sect.~\ref{subsec:Wavelet scattering transform}. The key innovation of our approach, presented in Sect.~\ref{subsec:WST $m$-mode ratios}, is a new statistical measure -- the WST $m$-mode ratios -- derived from the WST coefficients. The full analysis framework and its application to simulated datasets are described in the subsequent sections.

\subsection{3D Wavelet scattering transform}
\label{subsec:Wavelet scattering transform}
WST was originally developed in the context of signal processing as a structured compromise between traditional statistical measures and deep learning techniques such as convolutional neural networks (CNNs). It applies a sequence of fixed, non-learnable convolutional filters combined with nonlinear modulus and averaging operations. With analytically defined and interpretable components, WST efficiently captures non-Gaussian and higher-order statistical features, offering a robust and stable representation of complex data fields.

In WST, an input field  $I(\bm{x})$ is convolved with a set of localized wavelets generated through spatial and angular dilations of a mother wavelet, allowing the extraction of features at multiple scales and orientations. The convolved outputs are passed through a nonlinear modulus operator and then spatially averaged, yielding the WST coefficients that are sensitive to the clustering properties of the input field at specific scales and directions.

In this study, we adopt the 3D WST implementation introduced  in~\cite{valogiannis_going_2022, valogiannis_towards_2022}, and compute the WST coefficients up to the second order ($n=2$) using the open-source \texttt{kymatio} package\footnote{https://github.com/kymatio}~\citep{eickenberg_solid_2018}. When $\Psi_{j, l}(\bm{x})$ represents a wavelet at scale $j$ and orientation $l$, WST operation acts on the input field $I(\bm{x})$ as follows: 
\begin{equation}\label{eq:Ijl}
I_{j,l}(\bm{x}) = \left| I(\bm{x}) \otimes \Psi_{j, l}(\bm{x}) \right|\,,
\end{equation}
where $\otimes$ denotes convolution. 

A family of wavelets $\Psi_{j, l}^{m}(\bm{x}) $ is typically constructed by applying dilations and rotations to a mother wavelet, expressed as: 
\begin{equation}
\Psi_{j, l}^{m}(\bm{x}) = 2^{-3j}\Psi_l^m(2^{-j}\bm{x})\,,
\end{equation}
where $\Psi_l^m(\bm{x})$ is the mother wavelet, a solid harmonic wavelet multiplied by a Gaussian envelope, expressed as: 
\begin{equation}
\Psi_l^m(\bm{x}) = \frac{1}{(2\pi)^{3/2}} \exp\left(-\frac{|\bm{x}|^2}{2\sigma^2}\right) |\bm{x}|^l Y_l^m\left( \frac{\bm{x}}{|\bm{x}|} \right)\,,
\end{equation}
where $Y_l^m$ are the spherical harmonics, and $\sigma$ is the width of the Gaussian, set to $\sigma = 0.8$ in this study. 

By applying a set of localized wavelets $\Psi_{j, l}^m(\bm{x})$ across scales $j$ and orientations $l$, one obtains the WST coefficients $S_n$--real-valued quantities that characterize the field. The coefficients $S_n$ for $n = 0, 1$ and $2$ are defined as follows:

\begin{equation}
\label{eq:S_1}
\begin{aligned}
S_0 & = \left\langle |I_0(\bm{x})^q| \right\rangle\,,\\
S_1(j, l, m) & = \left\langle |I_0(\bm{x}) \otimes \Psi_{j,l}^{m}(\bm{x})| ^{q} \right\rangle\,, \\
S_2(j', j, l, m) & = \left\langle |U_1(\bm{x}) \otimes \Psi_{j',l}^{m}(\bm{x})| ^{q} \right\rangle\,,  
\end{aligned}
\end{equation}
with
\begin{equation}
U_1(j, l, m)(\bm{x}) = |I_0(\bm{x}) \otimes \Psi_{j,l}^{m}(\bm{x})|\,,
\end{equation}
where $\langle \cdot \rangle$ denotes spatial averaging, and $q$ is an exponent controlling the emphasis of field features: $q > 1$ enhances overdense regions, $q < 1$ emphasizes underdense ones, and $q = 1$ corresponds to the standard WST. In this study, we focus exclusively on the case $q = 1$.

Given a 3D input field and a set of spatial dyadic scales $J$ and orientations $L$, the WST coefficients can be computed for any order. Here, the indices are $j \in [0, 1, \dots, J]$, $l \in [0, 1, \dots, L]$, and $m \in [-l, \dots, 0, \dots, l]$, while $j' > j$, since $j' < j$ does not contribute any substantial cosmological information~\citep{cheng_new_2020}.
 
\begin{figure*}[htbp]
\centering
\includegraphics[width=0.9\textwidth]{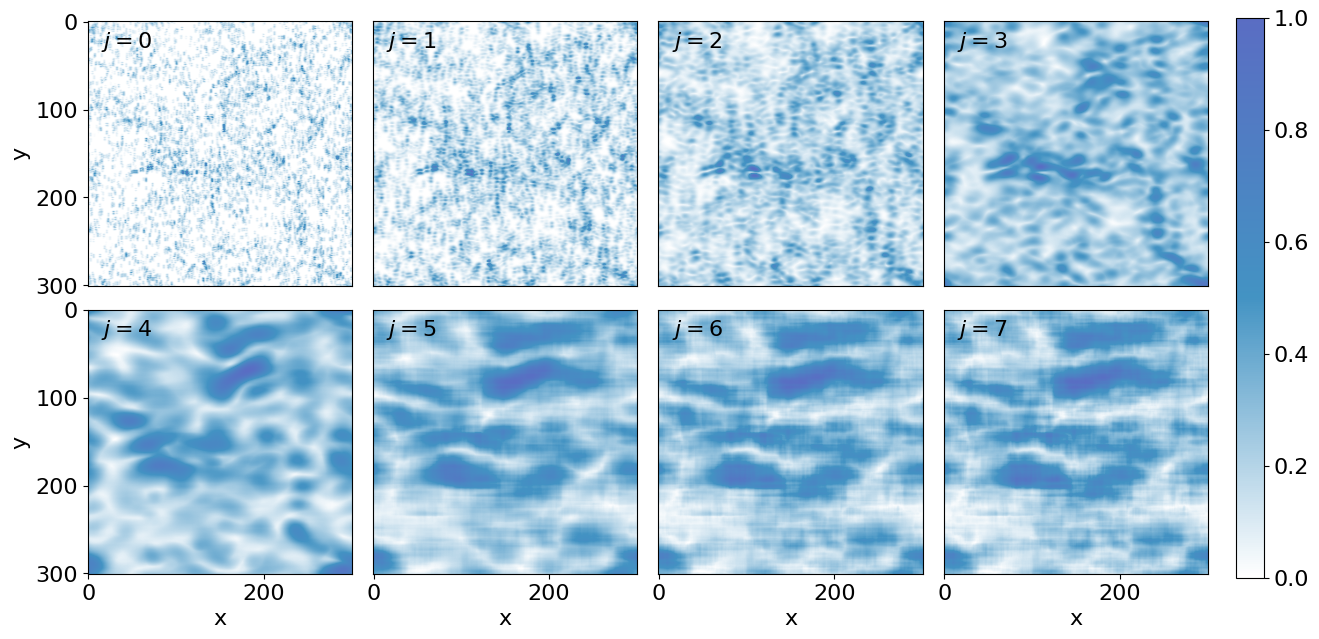}
\caption{Resulting 1D projection of first order WST component images (the same region from figure~\ref{fig:b}) are computed with $j$ ranging from 0 to 7 and fixed $l = 1$, $m=1$, for the  $\texttt{WMAP}$ dataset with the high-density apodization.  As $j$ increases, the results reveal progressively larger-scale structures, demonstrating that different $j$ values extract features at different spatial scales. Since the value range of each component varies significantly, each image is rescaled by min-max Normalization to make the values dimensionless. }
\label{fig:f}
\end{figure*}

To illustrate WST intuitively, Fig.~\ref{fig:f} shows the component images $I^m_{j,l}(\bm{x})$ (defined in Eq.~\ref{eq:Ijl}) for $j = 0$ to 7, with fixed $l = 1$ and $m = 1$, using \texttt{WMAP} data detailed in Sect.~\ref{sec:Simulation datasets}. As $j$ increases, the images reveal progressively broader features, demonstrating how varying $j$ isolates features at different spatial scales.

\subsection{WST $m$-mode ratios}
\label{subsec:WST $m$-mode ratios}
We introduce a novel statistical quantity, \textit{the WST $m$-mode ratios, $R^{\rm wst}(j,l)$}, designed to enhance sensitivity to cosmological parameters while reducing tracer-bias dependence. Alternative strategies--such as $l$-normalized quantity $S_{jl}/\langle S_{jl}\rangle_{l}$, $j$-$l$ normalized quantity $S_{jl}/\langle S_{jl}\rangle_{jl}$, power-law scattering transforms $S^\alpha$, and logarithmic scattering transforms $\log S$--either fail to sufficiently suppress bias sensitivity or diminish cosmological sensitivity. In contrast, $R^{\rm wst}(j,l)$ achieves both, making it the only approach that meets our requirements.

The $m$-mode ratios for $n=1$ and $n=2$ are defined, respectively, as
\begin{equation}\label{eq:R1}
R^{\mathrm{wst}}(j, l) = \frac{1}{l} \sum_{m=0}^{l-1} \frac{S_1(j,l,m+1)}{S_1(j,l,m)}\,, \quad l \neq 0\,,
\end{equation}
and 
\begin{equation}\label{eq:R2}
R^{\mathrm{wst}}_2(j', j, l) = \frac{1}{l} \sum_{m=0}^{l-1} \frac{S_2(j',j,l,m+1)}{S_2(j',j,l,m)}\,, \quad l \neq 0\,,
\end{equation}
where $S_1$ and $S_2$ are the WST coefficients. For a real field, these coefficients satisfy the symmetry relation $S^m = S^{-m}$, so we only consider the coefficients with $m \geq 0$, as no additional information is provided for $m < 0$. For $l = 0$, only the $m = 0$ mode exists, so no meaningful division of modes can be performed; thus, we exclude this case.

Physically, this ratio-based statistic isolates localized directional modulations in the field while largely canceling isotropic contributions from tracer bias. It characterizes the directional evolution of WST coefficients across azimuthal harmonic modes $m$, effectively capturing anisotropic signal features and remaining robust against global amplitude modulations induced by tracer bias. As shown in Appendix~\ref{app:bias}, these ratios are independent of linear bias for both 
$n=1$ and $n=2$. Appendix~\ref{app:dir} further demonstrates, via a physically motivated example, that at linear order the ratios depend solely on the strength of directional modulation. This confirms that the cancellation of isotropic bias contributions holds for both $n=1$ and  $n=2$.

We have computed the WST coefficients up to second order ($n=2$). Including second-order coefficients captures scale-to-scale correlations and interactions between wavelet responses, providing complementary information on non-Gaussianity. We find that the proposed $m$-mode ratios remain effective at $n=2$, continuing to suppress the impact of tracer bias. However, in terms of sensitivity to cosmological parameters and tracer bias, the gain from $n=2$ is limited. Considering the much higher computational cost and the limited gain, we focus on the first order coefficients in the main analysis and all $n=2$ results are presented in Appendix~\ref{app:s2} for completeness.

\section{Simulation datasets}
\label{sec:Simulation datasets}
Our analysis is based on the \texttt{CosmicGrowth} Simulations under a $\Lambda$CDM cosmology, a set of N-body cosmological simulations evolved with an adaptive parallel P$^3$M N-body code~\citep{jing_cosmicgrowth_2019}. We selected two simulations based on cosmological parameters from the WMAP~\citep{2013ApJS..208...19H} and Planck~\citep{Planck:2013pxb} missions, with the values (WMAP/Planck) given by: $\Omega_m = 0.268/0.315$ (where $\Omega_m = \Omega_b + \Omega_c$), $\Omega_b = 0.0445/0.0487$, $\sigma_8 = 0.83/0.829$, $n_s = 0.968/0.9603$, and $h = 0.71/0.673$.

The simulations we selected contain $2048^3$ DM particles within a comoving box with a side length of $1.2~{\rm Gpc}/h$. They are evolved from redshift $z = 72$ to $z = 0$ using identical initial conditions. The effect of redshift-space distortions is also accounted for in this study.  DM halos in each snapshot of the \texttt{CosmicGrowth} simulations were identified using the Friends-of-Friends (FoF) algorithm, with merger trees and subhalo catalogs subsequently constructed using the Hierarchical Branch Tracing (HBT) algorithm~\citep{HBT2012}.

This study is part of the preparation for upcoming CSST-like Stage-IV slitless spectroscopic galaxy surveys, which are expected to observe galaxies primarily at redshifts $z < 1$~\citep{Gong_2019}. As a representative case, we constructed halo catalogs in redshift space at $z = 0.59$. To match the expected target number density of $5 \times 10^{-4}~h^3{\rm Mpc}^{-3}$, a minimum halo mass threshold (\texttt{WMAP}/\texttt{Planck}) of $M_{\rm min} \simeq 6.4 \times 10^{12}~h^{-1}M_\odot/7.4 \times 10^{12}~h^{-1}M_\odot$ was imposed, consistent with current spectroscopic survey limits~\citep{Miao:2023umi}. The density fields were constructed from the halo catalogs by assigning halos to a $1200^3$ grid using the CIC interpolation scheme, resulting in a cell size of $1~h^{-3}{\rm Mpc}^3$. 

\subsection{Simulation purpose}
\label{subsec:Purpose of simulations}

We now emphasize that the two simulations based on WMAP and Planck cosmological parameters--resulting in three datasets, denoted as \texttt{mWMAP}, \texttt{WMAP} and \texttt{Planck}--are used for distinct purposes:

\begin{enumerate}
    \item \textbf{Sensitivity to cosmological parameters:} By comparing the WST $m$-mode ratios derived from \texttt{WMAP} and \texttt{Planck}, we can quantify how sensitive these coefficients are to differences in the underlying cosmological parameters. Both datasets are constructed to have the same number density, $n = 5 \times 10^{-4}~h^3{\rm Mpc}^{-3}$, with a mass cutoff of $M_{\rm cut}= 6.4 \times 10^{12}~h^{-1}M_\odot$ for \texttt{WMAP} and $ 7.4 \times 10^{12}~h^{-1}M_\odot$ for \texttt{Planck}. 
    
    \item \textbf{Sensitivity to tracer bias:}.
     Assuming that the tracer bias is primarily determined by halo mass at leading order--such that halos of different masses correspond to different bias values~\citep{Mo:1995cs,Sheth:1999su,Cooray:2002dia}--we generated two simulated datasets with distinct tracer biases based on the WMAP cosmology described above, denoted as \texttt{WMAP} and \texttt{mWMAP}. Specifically, a modified dataset \texttt{mWMAP} is created by selecting a different halo mass range, thereby modifying the tracer bias. By comparing the WST $m$-mode ratios derived from \texttt{WMAP} and \texttt{mWMAP}, we assess the sensitivity of these coefficients to changes in tracer bias. 
     
     In particular, \texttt{mWMAP} is constructed by shifting the halo mass range used in \texttt{WMAP} downward by 10\%, i.e.,
$M_{\rm halo} \in [6.4 \times 10^{12}, 1.0 \times 10^{15}]~M_\odot$ for \texttt{WMAP}, and $M_{\rm halo} \in [5.9 \times 10^{12}, 3.1 \times 10^{13}]~M_\odot$ for \texttt{mWMAP}.
\end{enumerate}

\subsection{High-density apodization}
\label{subsec:apodization}

We consider the following factors as significant concerns when analyzing the WST coefficients of cosmological density fields:

\begin{enumerate}
    \item \textbf{Correlation between halo mass/abundance and central density:} higher-mass halos tend to have denser centers, which can dominate the WST signal. If unaccounted for, this correlation may bias the interpretation of cosmological parameters, especially when comparing models with different halo mass distributions.
    
    \item \textbf{Smoothing effect of the WST convolution:} the WST convolution smooths local details, reducing sensitivity to low-density variations. This effect depends on the scale index $j$: stronger at high $j$ (large scales) and weaker at low $j$ (small scales), where variations are better captured.
    
    \item \textbf{Bias correlation with halo mass range and density distribution:} tracer bias is not uniform across all halos—it depends on both the halo mass range and the underlying density field. Cosmologies with predominantly high-mass halos have more concentrated central densities, while low-mass cosmologies exhibit smoother, sparser structures. Ignoring this can skew parameter inference.
    
    \item \textbf{Enhanced sensitivity of WST coefficients in high-density regions:} WST coefficients are more responsive to contrasts in high-density areas, which can amplify differences driven by tracer bias. This may lead to overestimating the influence of high-density structures in parameter estimation.
\end{enumerate}

In WST, the convolution smooths both clustering and structural information across scales. High-density central regions carry much more weight than surrounding low-density structures, causing clustering differences to dominate over structural information. This exaggerates tracer bias and makes cosmological parameter variations overly influenced by clustering. 

To mitigate these effects, we introduce a \textit{high-density apodization} using a tanh-based filter. This preprocessing technique gradually softens only the extremely high-density regions toward the global mean while leaving low-density structures unchanged. Physically, it preserves most structural information, reduces the dominance of extreme peaks, and ensures numerical stability by suppressing oscillations in WST space, similar to how sharp edges in real space produce oscillations in Fourier space. Several tests demonstrate that this method clearly outperforms alternative approaches (e.g., original density scheme, direct density-truncation, and mark-weighted density ~\citep{yang_using_2020}), offering strong suppression of tracer-bias effects while preserving cosmological parameter sensitivity. Its simplicity and robustness make it a practical solution for bias control.

Specifically, our proposed high-density apodization replaces high-density values with softened values that transition toward the global mean, using a tanh-based decay function defined as: 
\begin{equation}\label{eq:m}
\begin{aligned}
\rho_{\rm apo} (\bm{x})= & \frac{\rho(\bm{x})}{2} \left[1 + \tanh(\alpha)\right] + \frac{\bar{\rho}}{2} \left[1 - \tanh(\alpha)\right]\,, \\
&{\rm with}~~\alpha = k\big(\rho_\epsilon - \rho(\bm{x})\big)\,,
\end{aligned}
\end{equation}
where $\rho_{\rm apo}$ denotes the apodized density, $\rho_\epsilon$ is the threshold density above which apodization is applied, $\bar{\rho}$ is the global mean density of the field, and $k > 0$ controls the sharpness of the transition. The resulting $\rho_{\rm apo}$ is then transformed into the overdensity field $\delta$ for further analysis. Specifically, we sort the density values of each pixel in descending order and define $\rho_\epsilon$ as the density value at the top 15\%. The resulting thresholds are $\rho_\epsilon =  3.9 \times 10^{12}$, $2.9 \times 10^{12}$, and $4.6 \times 10^{12}$ $({\rm Mpc}/h)^{-3}$, and $\bar{\rho}= 1.0 \times 10^{10}$, $8.5 \times 10^{9}$, and $5.7 \times 10^{9}$ $({\rm Mpc}/h)^{-3}$ for \texttt{WMAP}, \texttt{mWMAP}, and \texttt{Planck}, respectively.  We set $k=0.4$ to control the sharpness of the transition.

% In WST, the convolution smooths both clustering and structural information across scales. High-density central regions carry much more weight than surrounding low-density structures, causing clustering differences to dominate over structural information. This exaggerates tracer bias differences and makes cosmological parameter variations overly influenced by clustering. High-density apodization (tanh-based smoothing) mitigates this effect while preserving low-density structures. This softening is therefore necessary to reduce bias and maintain sensitivity to structure, and its simplicity and effectiveness make it a practical choice.

\begin{figure}[htbp]
% \label{fig:g}
\centering
\includegraphics[width=0.45\textwidth]{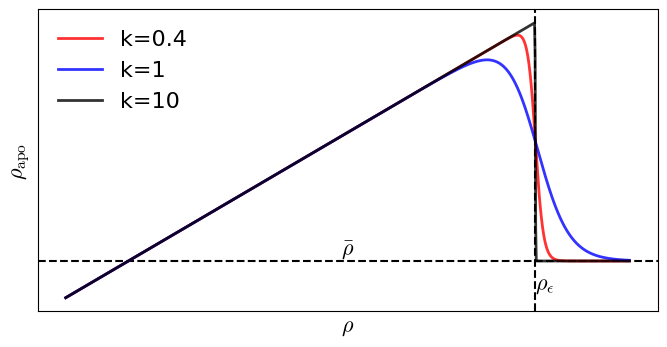}
\caption{Tanh-based decay profiles for different values of $k$. We adopt $k=0.4$ in this study.}
\label{fig:g}
\end{figure}

As shown in Fig.~\ref{fig:g}, the tanh-based decay functions are plotted for different values of $k$. As observed, when $\rho < \rho_\epsilon$, $\rho_{\rm apo} \approx \rho$, and when $\rho > \rho_\epsilon$, $\rho_{\rm apo} \approx \bar{\rho}$. A sharp transition in the shape occurs near $\rho \approx \rho_\epsilon$. Larger values of $k$ result in sharper transitions at the threshold. This high-density apodization causes high-density regions to converge toward the global mean density.

We then applied this filtering scheme to \texttt{mWMAP}, \texttt{WMAP}, and \texttt{Planck} data, in order to suppress the contributions from extremely high-density regions. These apodized density fields were subsequently converted into three-dimensional (3D) matter overdensity fields, $ \delta(\bm{x})$, defined as $\delta(\bm{x}) = \rho(\bm{x})/\bar{\rho} - 1$. WST was then applied to these 3D overdensity fields.

\begin{figure}[htbp]
\centering
\includegraphics[width=.5\textwidth]{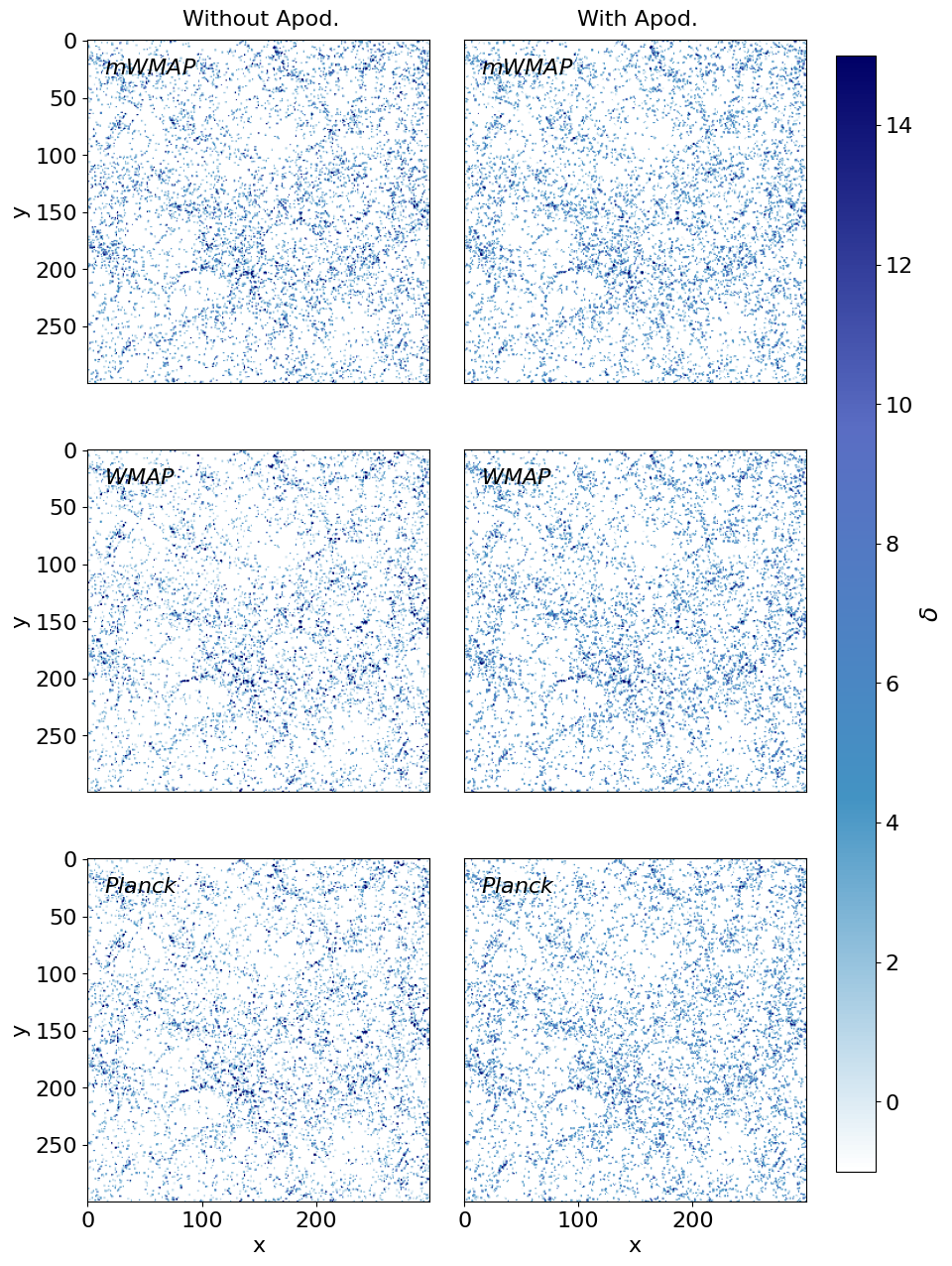}
\caption{A comparison of the overdensity fields before (left) and after (right) applying high-density apodization across different datasets. Each panel shows a 1D projection of the field, computed on a $300^2$ grid spanning 300 ${\rm Mpc}/h$ per side, with a slice thickness of  $150~{\rm Mpc}/h$. The first, second, and third rows show slices from the \texttt{mWMAP}, \texttt{WMAP}, and \texttt{Planck} datasets, respectively. As observed, the apodization scheme suppresses high-density peaks by shifting them toward the mean density, while leaving low-density regions largely unchanged.  }
\label{fig:b}
\end{figure}

Fig.~\ref{fig:b} shows a comparison of the overdensity field before and after applying high-density apodization, across different datasets. Each panel presents a 1D projection of the overdensity field, computed on a grid of $300 \times 300$ grids spanning 300 ${\rm Mpc}/h$ per side. The thickness of each slice along the projection axis is 150 ${\rm Mpc}/h$. As seen, the apodization scheme effectively suppresses the high-density regions, shifting them toward the mean density, while leaving the low-density regions almost unchanged.

\section{chi-square statistic}
\label{sec:chi-square statistic}
We use the $\chi^2$ statistic to quantify the sensitivity of the proposed measure--{\it WST $m$-mode ratios}--to cosmological parameters and tracer bias. Let us first introduce the general form of the $\chi^2$ statistic. Given a set of model-predicted observables represented as a vector $\bm{p}^{\rm model}(\theta)$, and a corresponding set of observed data $\bm{p}^{\rm data}$, one aims to minimize the $\chi^2$ value to obtain the best-fit estimate of the model parameters $\theta$. If the model parameters deviate from their true values, the predicted observables can differ significantly from the data, resulting in a large $\chi^2$ value. This makes the $\chi^2$ statistic highly sensitive to changes in $\theta$. Throughout this study, we use $\bm{p}$ to denote the vector of collected $R^{\rm wst}$ for the $m$-mode ratio analysis. For comparison of sensitivities, we also use $\bm{p}$ to represent the corresponding vectors of other statistical measures, as listed in Sect.~\ref{sect:other}. 

The $\chi^2$ statistic is given by:
\begin{equation}\label{eq:chi2}
\chi^2= (\bm{p}^{\rm data} - \bm{p}^{\rm model})^T \mathbf{Cov}^{-1} (\bm{p}^{\rm data} - \bm{p}^{\rm model})\,,
\end{equation}
where the covariance matrix $\mathbf{Cov}$ is usually estimated empirically from a large ensemble of simulations that are designed to statistically match the data based on a fiducial cosmology.  

In our analysis, we compute the $\chi^2_{\rm cos}$ statistic using the \texttt{WMAP} and \texttt{Planck} datasets to assess sensitivity to cosmological parameters, and compute $\chi^2_{\rm bias}$ using \texttt{WMAP} and \texttt{mWMAP} to evaluate sensitivity to tracer bias. In this context, we treat the observables from \texttt{WMAP} as the data vector $\bm{p}^{\rm data}$,  while the observables from \texttt{Planck} and \texttt{mWMAP} serve as the model prediction $\bm{p}^{\rm model}$. Specifically, the statistic $\chi^2_{\rm cos}$ is defined as
\begin{equation}
\chi^2_{\rm cos} = (\bm{p}^{\rm WMAP} - \bm{p}^{\rm Planck})^T \mathbf{Cov}^{-1} (\bm{p}^{\rm WMAP} - \bm{p}^{\rm Planck})\,,
\end{equation}
and the sensitivity to tracer bias is quantified by
\begin{equation}
\chi^2_{\rm bias} = (\bm{p}^{\rm WMAP} - \bm{p}^{\rm mWMAP})^T \mathbf{Cov}^{-1} (\bm{p}^{\rm WMAP} - \bm{p}^{\rm mWMAP})\,.
\end{equation}

We assume that the dataset \texttt{WMAP} represents the fiducial cosmology, and we estimate the covariance matrix based on this dataset. Since we only have a single simulation, we divide it uniformly into $6^3$ sub-boxes, each with a side length of $200~{\rm Mpc}/h$, to estimate the covariance. The covariance matrix estimated from the sub-boxes is given by:

\begin{equation}
\mathbf{Cov}_{\rm sub} = \frac{1}{N-1} \sum_{i=1}^{N} \left( \Delta \bm{p}_i \right) \left( \Delta \bm{p}_i \right)^{T}\,,
\end{equation}

where $\Delta \bm{p}_i = \bm{p}_i - \langle \bm{p} \rangle$ represents the deviation of the $i$-th sub-box from the mean of all sub-boxes. In an ideal scenario with many independent realizations, the true covariance of the large box would correspond to the expectation of the covariance matrix averaged over those realizations. However, since only a single realization is available, the variance of the large box is smaller than that of the fluctuations in the sub-boxes.  For independent samples, the variance of the sample mean scales as $1/N$. Therefore, the true covariance of the large box is estimated by rescaling the covariance matrix $\mathbf{Cov}_{\rm sub}$ by a factor of $1/N$. The corrected covariance estimate for the large box is therefore given by:
\begin{equation}
\mathbf{Cov} \approx \frac{1}{N} \mathbf{Cov}_{\rm sub}\,.
\end{equation}

In addition, to account for the varying number of statistical measures used across different $j$ ranges, we consider the {\it reduced chi-squared} statistic, defined as
\begin{equation}
    \chi^2_{\nu} = \frac{\chi^2}{N_\nu}\,, \quad \text{with} \quad N_\nu = N_{\rm data} - N_{\rm param}\,,
\end{equation}
where $N_\nu$ is the number of degrees of freedom. This normalization enables a fair comparison between cases with different amounts of data or numbers of model parameters. In our case, since we do not perform parameter fitting, the degrees of freedom are given by $N_{\nu} = N_{\rm data}$.

For a statistically consistent model and correctly estimated uncertainties, the reduced chi-squared is expected to lie within the $1\sigma$ range of
\begin{equation}\label{eq:chi2_sigma}
    \chi^2_{\nu} = 1 \pm \sigma_\nu\,, \quad \text{with}~~\sigma_\nu=\sqrt{\frac{2}{N_\nu}}\,.
\end{equation}
A value of $\chi^2_{\nu} \approx 1$ indicates a good agreement between the model and the data, while significant deviations from this range may signal overfitting, model misspecification, or misestimated uncertainties. In this sense, $\chi^2_{\nu} $ quantifies the sensitivity to discrepancies between the predicted and observed statistics.

\section{Results}\label{sec:results}

In this section, we present the resulting $\chi^2$ values and justify the preferred $j$ range for $R^{\rm wst}$. We also compare its performance with the original WST coefficients and two-point statistics, namely the power spectrum and the normalized power spectrum. We find that $j \in [3,7]$ and $l \in [1,2]$ effectively minimize tracer bias while retaining high sensitivity to cosmological parameters. This selection corresponds to ten coefficients: for each of the five $j$ values, we include $R^{\rm wst}(j, l=1)$ and $R^{\rm wst}(j, l=2)$, giving a total of $5 \times 2 = 10$ coefficients.

\subsection{Sensitivities of WST $m$-mode ratios}\label{sec:R}

\begin{figure*}[htbp]
\centering
\includegraphics[width=1.0\textwidth]{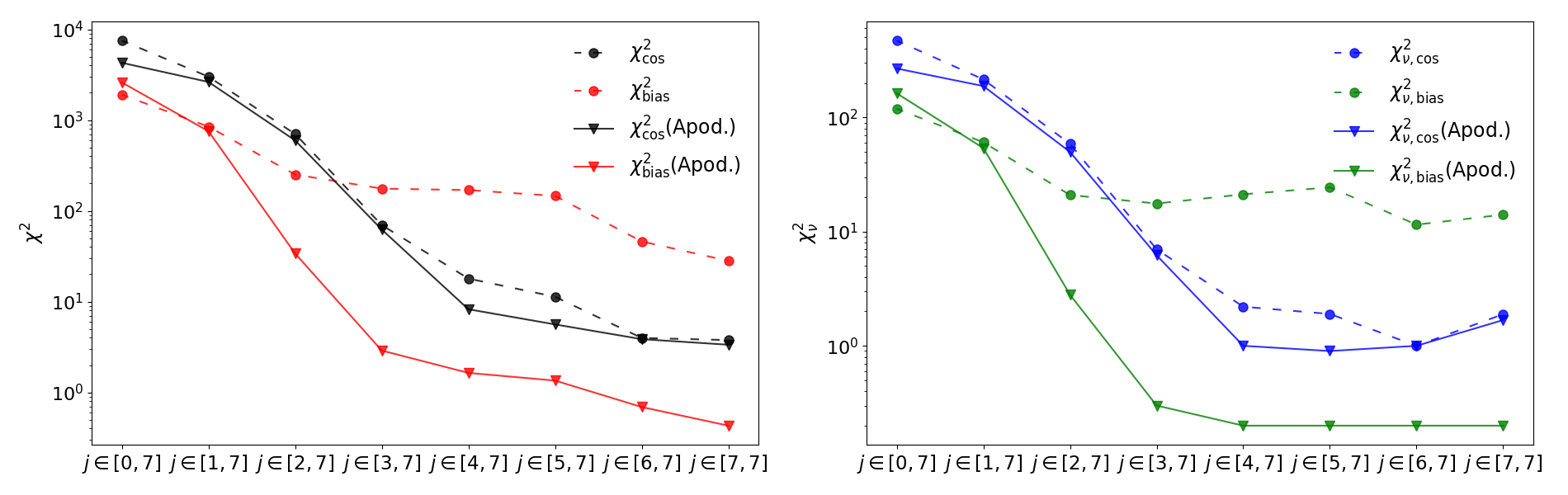}
\caption{ Results for $\chi^2$ (left) and $\chi^2_{\nu}$ (right), with eight cases corresponding to different ranges of $j$, while keeping $l \in [1,2]$ fixed. The results for the $j$ ranges from $[0,7]$ to $[7,7]$ are shown respectively, corresponding to different $R^{\mathrm {wst}}$ measurements, leading to different $\chi^2$ and $\chi^2_{\nu}$ values. Each panel compares the results with (triangle) and without (solid dot) the high-density apodization scheme.}
\label{fig:chi2}
\end{figure*}

In Fig.~\ref{fig:chi2}, we present our results for $\chi^2$ (left) and the reduced chi-squared $\chi^2_{\nu}$ (right), respectively. In each panel, both  $\chi_{\rm cos}$ and $\chi_{\rm bias}$ are shown to illustrate the sensitivities of $R^{\rm wst}$, with comparisons made between results with and without the high-density apodization scheme.

We consider eight cases with varying ranges of $j$, while keeping $l\in [1,2]$ fixed. Varying the range of $j$ changes the number of $R^{\rm wst}$ measurements, which in turn affects the covariance estimation. Specifically, the choices $j \in [0, 7]$, $[1, 7]$, $[2, 7]$, $\dots$, $[7, 7]$ correspond to $N_\nu = 16, 14, 12, \dots, 2$ degrees of freedom, respectively--each representing measurements of $R^{\rm wst}$ with progressively fewer modes included. The associated uncertainties are given by Eq.~\ref{eq:chi2_sigma}, resulting in $\sigma_\nu \approx 0.35$, 0.38, 0.41, 0.45, 0.50, 0.58, 0.71 and 1.0, respectively.

From the left panel, we observe that $\chi^2_{\rm cos}$ decreases rapidly as the range of $J$ becomes narrower. This is because using the full range of $J$ captures more cosmological information, resulting in higher sensitivity. Additionally, applying high-density apodization yields lower $\chi^2$ values compared to the case without apodization, as it smooths out high-density regions, which may slightly reduce sensitivity to cosmological parameters. These trends hold true in the right panel as well, where the number of degrees of freedom is properly accounted for.

The overall $\chi^2_{\nu}$ and $\chi^2$ values for each case are summarized in Tab.~\ref{tab:chi2}. Based on the amplitude of $\chi^2_{\rm cos}$ alone, one may opt for $j \in [0,7]$ without apodization, as it appears optimal for distinguishing cosmological parameters. However, the corresponding $\chi^2_{\rm bias}$ is large, indicating strong sensitivity to the tracer bias. Notably, $\chi^2_{\rm bias}$ decreases sharply as the $j$ range narrows from $[0,7]$ to $[3,7]$. The most remarkable feature is that applying apodization exponentially suppresses $\chi^2_{\rm bias}$--by about three orders of magnitude. For even narrower $j$ ranges, the $\chi^2_{\rm bias}$ values converge to a small value around 1. In contrast, without apodization, the decrease in $\chi^2_{\rm bias}$ is modest, dropping only from 175.8 to 28.15 as the range changes from $[3,7]$ to $[7,7]$. 

Moreover, in the case of using the high-density apodization, $\chi^2_{\nu, \rm bias}$ falls within the range of 0.2--0.3 for $j \in [3, 7]$, $[4, 7]$, $\dots$, $[7, 7]$, deviating from one (the expected value of $\chi_{\nu}$) by less than $1\sigma$. However, for $j \in [2, 7]$, $[1, 7]$, and $[0, 7]$, the deviation quickly grows to more than $4\sigma$ (e.g., $\chi_{\nu, \rm bias} = 2.8$ for $j \in [2, 7]$, with $\sigma_\nu = 0.41$). These results demonstrate that narrower $j$-ranges, such as those smaller than $[3, 7]$, are less sensitive to the trace bias.

For $j \in [3, 7]$, $\chi_{\nu, \rm cos} = 6.2$ and $7.0$ for the cases without and with apodization, respectively, both showing a deviation of more than $10\sigma$ from one, indicating strong sensitivity to distinguishing a specific cosmological parameter range. In contrast, the values are about $1.7\sigma$ consistent with each other, suggesting only a slight loss of sensitivity when using apodization.

Although the wider $j$-ranges, from $[0, 7]$ to $[2, 7]$, lead to very large $\chi_{\nu, \rm cos}$ values in the range of 267.6 to 49.8, indicating higher sensitivity to cosmological parameters, the sensitivity to the tracer bias also increases significantly. This is a scenario we want to avoid. Combining these findings, we adopt $j \in [3,7]$ as the optimal choice, providing a good compromise between high sensitivity to cosmological parameters and low sensitivity to the tracer bias.

\begin{table*}
\centering 
\caption{Reduced chi-square ($\chi^2_\nu$) and total chi-square ($\chi^2$) values for $R^{\rm wst}$ quantifying the sensitivity to cosmological parameters and tracer bias are evaluated for different $j$ ranges, with adopting two $l$ values $(l= 1, 2)$, both with and without the high-density apodization.  The $1\sigma$ uncertainties, $\sigma_\nu$, derived from the number of degrees of freedom for $\chi_\nu$, are also listed in the last row for comparison.
}\label{tab:chi2} 
\resizebox{\textwidth}{!}{
\begin{tabular}{llcccccccc}
\hline
\hline
\textbf{$R^{\rm wst}$--Parameter} & \textbf{Apod.} & $j\in[0,7]$  & $[1,7]$ & $[2,7]$ & $[3,7]$ & $[4,7]$ & $[5,7]$ & $[6,7]$ & $[7,7]$ \\
\hline
\multirow{2}{*}{$\chi^2_{\nu,\mathrm{cos}}$/$\chi^2_{\mathrm{cos}}$} 
    & without & 469.4/7509.7 & 215.0/3009.5 & 58.6/702.6 & 7.0/69.7 & 2.2/18.0 & 1.9/11.3 & 1.0/4.0 & 1.89/3.77 \\
    & with    & 267.6/4281.3 & 187.9/2630.9 & 49.8/597.2 & 6.2/62.2 & 1.0/8.2 & 0.9/5.6 & 1.0/3.9 & 1.68/3.35 \\ 
\hline
\multirow{2}{*}{$\chi^2_{\nu,\mathrm{bias}}$/$\chi^2_{\mathrm{bias}}$} 
    & without & 118.9/1902.8  & 60.5/847.0 & 20.9/251.2 & 17.6/175.8 & 21.2/169.5 & 24.4/146.1 & 11.5/46.1& 14.07/28.15 \\
    & with    & 162.4/2598.5 & 53.8/753.6 & 2.8/33.9 & 0.3/2.9 & 0.2/1.6 & 0.2/1.4 & 0.2/0.7& 0.22/0.43\\
\hline

\multirow{1}{*}{$\sigma_{\nu}$} 
    & / & 0.35  & 0.38 &  0.41 &  0.45 & 0.50 & 0.58 & 0.71 & 1.0\\
\hline
\hline
\end{tabular}
}
\end{table*}

\subsection{Other Statistical measures}\label{sect:other}

To compare the sensitivities of the WST $m$-mode ratios with those of the original WST coefficients and the commonly used statistic, the power spectrum, we also analyzed the sensitivity of the WST coefficients averaged over $m$ to both the cosmological parameters and the tracer bias. The $m$-averaged WST coefficients for $S_1$ (as defined in Eq.~\ref{eq:S_1}) are denoted as $\overline{S}_1$, and are expressed as:
\begin{equation}
\overline{S}_1(j, l) = \frac{1}{2l + 1} \sum\limits_{m} S_1(j, l, m)\,.
\end{equation}
Due to the symmetry, we have $S_1(j, l, m) = S_1(j, l, -m)$, which implies that negative values of $m$ do not contribute any additional information to the analysis. We therefore limit the values to $m=0, 1, 2$ when calculating the $\chi^2$ and $\chi^2_{\nu}$ values.

Moreover, we will explore other statistical measures, including the power spectrum and its variant, the normalized power spectrum, to evaluate their sensitivity performance. In practice, the power spectrum is computed directly from the Fourier-transformed density field as
\begin{equation}\label{eq:pk}
P(\bm{k}) = |\tilde{\delta}(\bm{k})|^2 \,,
\end{equation}
where $\tilde{\delta}(\bm{k})$ is the Fourier transform of the overdensity field $\delta(\bm{x})$. 
The spherically averaged, one-dimensional power spectrum $P(k)$ is then obtained by averaging $P(\bm{k})$ over all directions for a fixed $k$. We adopt a linear binning scheme, with $k \in [0.045, 0.135]$ divided into 10 bins.

As an alternative, we consider a scale-normalized power spectrum to test its sensitivity. We define the dimensionless quantity $k^3 P(k)$ to highlight scale-dependent features and normalize it within each $k$-bin by its mean:
\begin{equation}
P_{\rm norm}(k) = \frac{k^3 P(k)}{\langle k^3 P(k) \rangle_i}\,,
\end{equation}
for the $i$-th bin. This normalization emphasizes relative mode-to-mode variations rather than absolute amplitudes, mitigating the impact of tracer bias while reducing sensitivity to cosmological parameters. Since our analysis is based on $m$-mode ratios, we primarily adopt this normalized form for comparison with WST ratios.

\subsection{Sensitivity performance comparison}

\begin{table*}
\centering 
\caption{Reduced chi-square ($\chi^2_\nu$) and total chi-square ($\chi^2$) values for $m$-averaged WST coefficients $\overline{S}_1$, quantifying the sensitivity to cosmological parameters and tracer bias, are evaluated for different $j$-ranges, with and without the high-density apodization. Three values of $l$ ($l = 0, 1, 2$) are used for each $j$-range for $\overline{S}_1$. The $1\sigma$ uncertainties for $\chi_\nu$, $\sigma_\nu$, derived from the degrees of freedom for each case, are listed in the last row for comparison.
}\label{tab:chi2_S} 
\resizebox{\textwidth}{!}{
\begin{tabular}{llcccccccc}
\hline
\hline  
\textbf{$\overline{S}_1$--Parameter} & \textbf{Apod.} & $j\in[0,7]$  & $[1,7]$ & $[2,7]$ & $[3,7]$ & $[4,7]$ & $[5,7]$ & $[6,7]$ & $[7,7]$ \\
\hline     
\multirow{2}{*} {$\chi^2_{\nu,\mathrm{cos}}$/$\chi^2_{\mathrm{cos}}$} 
    & without & 58.9 / 1413.2 & 63.5 / 1333.4 & 45.7 / 823.5 & 33.2 / 498.5 & 22.1 / 264.9 & 18.2 / 164.1 & 17.6 / 105.8 & 30.2 / 90.5 \\
    & with    & 196.1 / 4707.5 & 203.2 / 4266.6 & 92.6 / 1666.0 & 59.6 / 893.4 & 33.7 / 404.0 & 32.6 / 293.6 & 33.8 / 203.0 & 60.8 / 182.3 \\ 
\hline
\multirow{2}{*}{$\chi^2_{\nu,\mathrm{bias}}$/$\chi^2_{\mathrm{bias}}$} 
    & without & 996.1 / 23907.5 & 929.2 / 19512.7  & 966.1 / 17389.7 & 624.1 / 9361.4 & 399.8 / 4798.1 & 262.2 / 2360.0 & 248.9 / 1493.3 & 339.0 / 1017.1 \\
    & with    & 350.1 / 8401.6 & 117.7 / 2470.8 & 39.7 / 714.6 & 25.1 / 376.5 & 12.4 / 148.7 & 11.5 / 103.2 & 9.8 / 58.5 & 16.9 / 50.8\\
\hline
\multirow{1}{*}{$\sigma_{\nu}$} 
    & / & 0.29  & 0.31 &  0.33 &  0.36 & 0.41 & 0.47 & 0.58 & 0.82\\
\hline
\hline
\end{tabular}
}
\end{table*}

Tab.~\ref{tab:chi2_S} presents the $\chi^2_\nu$ and $\chi^2$ values for the $m$-averaged WST coefficients, $\overline{S}_1$, evaluated over various $j$-ranges ($j \in [0,7]$ to $[7,7]$), both with and without high-density apodization. For cosmological sensitivity ($\chi^2_{\nu, {\rm cos}}$), the inclusion of apodization increases the $\chi^2$ values across all $j$-ranges, indicating a emphasis of sensitivity to cosmological parameters. The $\chi^2_{\nu, {\rm cos}}$ values remain significantly larger than the expected statistical uncertainty $\sigma_\nu$ in all cases, indicating that the cosmological sensitivity of $\overline{S}_1$  is strong. Even in the most conservative scenario--$j \in [6,7]$ without apodization--the value of $\chi^2_{\nu, {\rm cos}} = 17.6$ still exceeds $\sigma_\nu = 0.58$ by a substantial amount.

In contrast, for the tracer bias sensitivity, applying apodization significantly lowers the $\chi^2_{\nu, {\rm bias}}$ values, indicating enhanced robustness to bias contamination. Among all cases, the range $j \in [6,7]$ with apodization achieves the lowest sensitivity to the tracer bias ($\chi^2_{\nu, {\rm bias}} = 9.8$). However, this value still greatly exceeds the expected statistical uncertainty ($\sigma_\nu = 0.58$), suggesting that the $\overline{S}_1$ measure may be less effective than the $R^{\rm wst}$ metric in bias insensitivity.

\begin{table}
\centering 
\caption{Same as Table~\ref{tab:chi2}, but showing the results for the power spectrum and normalized power spectrum, based on 10 $k$-bins in the range of $k\in[0.046,0.135]~h/{\rm Mpc}$.}
\label{tab:chi2_pk} 
\begin{tabular}{lccc}
\hline
\hline
\textbf{$P(k)$--Parameter} & \textbf{Apod.} & \textbf{$P(k)$} & \textbf{$P_{\rm norm}(k)$} \\
\hline
$\chi^2_{\nu,\mathrm{cos}}$/$\chi^2_{\mathrm{cos}}$ 
    & without & 37.8/377.63 & 0.2/1.6\\
    & with    & 17.5/175.0 & 0.2/2.1 \\
\hline
$\chi^2_{\nu,\mathrm{bias}}$/$\chi^2_{\mathrm{bias}}$ 
    & without & 354.0/3540.6 & 0.3/2.7 \\
    & with    & 4.8/47.6 & 0.04/0.4 \\
\hline
$\sigma_\nu$ 
    & /  & 0.45 & 0.45 \\
\hline
\hline
\end{tabular}
\end{table}

Moreover, Tab.~\ref{tab:chi2_pk} presents the reduced and total chi-square values for both the power spectrum $P(k)$ and the normalized power spectrum $P_{\rm norm}(k)$, evaluated over 10 $k$-bins in the range $k \in [0.046, 0.135]~h/{\rm Mpc}$, with and without high-density apodization. For cosmological sensitivity ($\chi^2_{\nu, {\rm cos}}$), the $P(k)$ values yield relatively small values compared to those without apodization ($\chi^2_{\nu, \mathrm{cos}} = 37.8$), which are reduced to $17.5$ with apodization. In contrast, the $P_{\rm norm}(k)$ values show negligible sensitivity in both cases ($\chi^2_{\nu, {\rm cos}} \approx 0.2$), indicating a significant loss of cosmological information due to the normalization.

Regarding the tracer bias sensitivity ($\chi^2_{\nu, {\rm bias}}$), apodization substantially reduces the values for both $P(k)$ and $P_{\rm norm}(k)$, with the reduced chi-square dropping from 354.0 to 4.8 for $P(k)$ and from 0.3 to 0.04 for $P_{\rm norm}(k)$, indicating strong suppression of bias contamination. However, in the case of $P(k)$, the reduced chi-square value of 4.8 remains significantly above the statistical uncertainty $\sigma_\nu = 0.45$, suggesting residual sensitivity to bias. In contrast, $P_{\rm norm}(k)$ yields extremely low $\chi^2$ values, consistent with the expected value of unity within $1\sigma$, but this comes at the cost of a severe loss of cosmological information. This trade-off highlights the limitation of $P_{\rm norm}(k)$ in capturing cosmological sensitivity. Overall, power spectrum-based statistics, including $P(k)$ and $P_{\rm norm}(k)$, appear less effective than the $R^{\rm wst}$ measure in jointly addressing bias insensitivity and cosmological discrimination.

\begin{figure*}[htbp]
\centering
\includegraphics[width=0.9\textwidth]{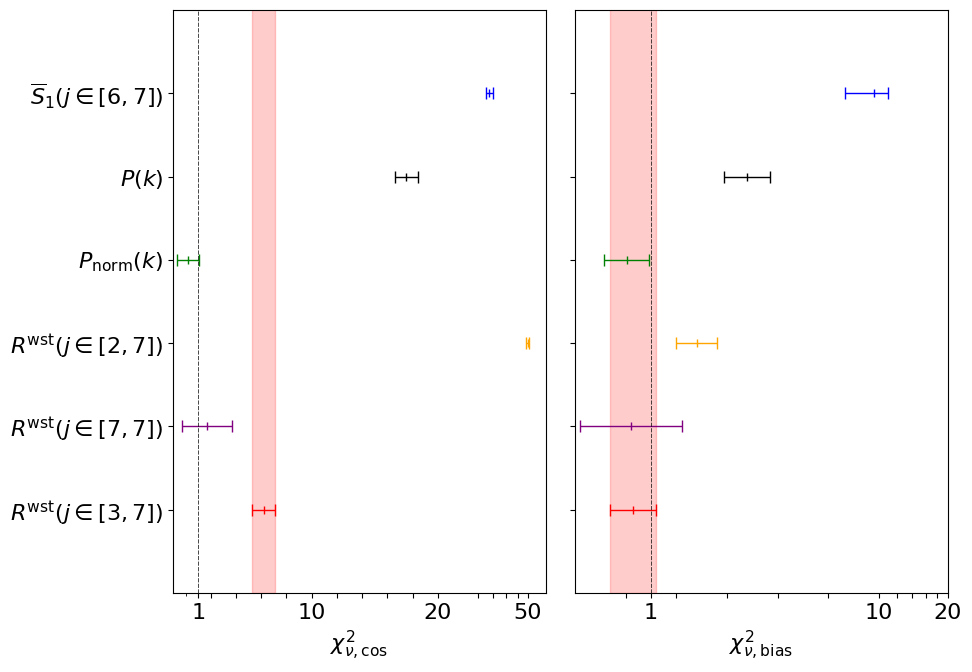}
\caption{Reduced chi-squared values, $\chi^2_\nu$, for cosmological parameters (left) and tracer bias (right), evaluated using various statistical summaries: WST $m$-mode ratios, $m$-averaged WST coefficients, the power spectrum, and the normalized power spectrum, all with the high-density apodization. The dotted line marks $\chi^2_\nu = 1$ (ideal fit), and each measurement includes its associated $2\sigma$ error bar. The red shaded regions indicate the chosen $R^{\rm wst}$ for $j \in [3,7]$, which offer high cosmological sensitivity with reduced tracer bias dependence. }\label{fig:chi2-all}
\end{figure*}

Fig.~\ref{fig:chi2-all} shows the values of the reduced chi-squared statistic, $\chi^2_{\nu}$, for the cosmological parameters (left) and tracer bias (right), evaluated using various statistical measures: $R^{\rm wst}$, $\overline{S}_1$, $P(k)$, and $P_{\rm norm}(k)$. This provides a comprehensive comparison of sensitivity and robustness. For all cases, the high-density apodization scheme is adopted. In both panels, the dotted vertical line marks $\chi^2_\nu = 1$, the expected value under an ideal model with correctly estimated uncertainties. The corresponding $2\sigma$ uncertainty, $2\sigma_\nu$, for each case is also indicated, as defined in Eq.~\ref{eq:chi2_sigma}. The red shaded bands highlight the $R^{\rm wst}$ results with $j \in [3,7]$, which are considered optimal.

Among all cases, $R^{\rm wst}$ offers the most favorable balance between high cosmological sensitivity and effective bias suppression. Specifically, $R^{\rm wst}(j \in [3,7])$ yields a reduced chi-square value of $\chi^2_{\nu, \rm cos} =6.2$, indicating strong sensitivity to the cosmological parameters, while simultaneously maintaining $\chi^2_{\nu, \rm bias} = 0.3$, which agrees well with the expected level of statistical fluctuations. For other $j$-ranges, such as $j \in [7,7]$, the tracer bias contamination remains modest ($\chi^2_{\nu, \rm bias} \lesssim 0.3$), but the cosmological sensitivity is significantly reduced, with $\chi^2_{\nu, \rm cos} \approx 2$, approaching the noise level. In contrast, the range $j \in [2,7]$ yields a large cosmological sensitivity ($\chi^2_{\nu, \rm cos} = 49.8$), but also shows a pronounced bias contamination ($\chi^2_{\nu, \rm bias} = 2.8$), which exceeds the $4\sigma$ statistical uncertainty ($\sigma_\nu = 0.58$).

This performance is significantly better than that of $\overline{S}_1 (j \in [6,7])$, which, despite offering some cosmological sensitivity, suffers from a larger bias sensitivity ($\chi^2_{\nu, \rm bias} =9.8$). Meanwhile, $P_{\rm norm}(k)$ offers strong suppression of bias-related effects; however, it captures almost no cosmological signal, as indicated by the very low value of $\chi^2_{\nu, \rm cos} \sim 0.2$. This demonstrates its limited applicability for cosmological analysis. 

In summary, only $R^{\rm WST}$ with $j \in [3,7]$ achieves both $\chi^2_{\nu, \rm cos} \approx 6$ and $\chi^2_{\nu, \rm bias} \sim 1$--a regime not reached by any other measure--highlighting its advantage in simultaneously capturing cosmological signals and controlling for tracer bias.

\section{Conclusions}
\label{sec:conclusion}

LSS in galaxy surveys has become a key probe for understanding the history of the universe. However, the presence of tracer bias between galaxies and halos poses a significant challenge in modern cosmology. WST offers a promising approach to extract rich statistical information from LSS. Yet, the sensitivity of WST to tracer bias remains largely unquantified. To address this, we introduce a new statistical framework based on WST that can distinguish LSS across different cosmological models while mitigating the impact of tracer bias. At the core of our method is the WST $m$-mode ratios, a novel and robust statistical measure. Furthermore, we develop a high-density apodization technique that reduces the influence of extreme values by smoothly rescaling them toward the global mean using a tanh-based decay function.

Using the \texttt{CosmicGrowth} N-body simulations, we analyze three datasets--\texttt{WMAP}, \texttt{mWMAP}, and \texttt{Planck}--to evaluate the performance of different measures in extracting cosmological information. The reduced chi-square statistics ($\chi^2_\nu$) show that, when combined with the apodization scheme, the WST $m$-mode ratios maintain strong sensitivity to cosmological parameters while remaining stable against halo mass selection effects.

Among all tested cases, $R^{\rm wst}$ with $j \in [3,7]$ provides the most favorable balance between cosmological sensitivity and bias control. This setting yields a high cosmological signal detection significance ($\chi^2_{\nu, \rm cos} = 6.2$) alongside a low bias contamination level ($\chi^2_{\nu, \rm bias} = 0.3$), consistent with expected statistical fluctuations.

Alternative scale ranges exhibit less optimal trade-offs. For instance, \textbf{$j \in [7,7]$} achieves modest bias suppression ($\chi^2_{\nu, \rm bias} \lesssim 0.3$) but suffers from reduced cosmological sensitivity ($\chi^2_{\nu, \rm cos} \approx 2$), approaching the noise floor. On the other hand, \textbf{$j \in [2,7]$} enhances cosmological sensitivity ($\chi^2_{\nu, \rm cos} = 49.8$) but incurs significant bias contamination ($\chi^2_{\nu, \rm bias} = 2.8$), exceeding the $4\sigma$ statistical threshold ($\sigma_\nu = 0.58$).

Furthermore, an exploration using $R_2^{\rm wst}$ showed consistent performance of $R^{\rm wst}$, whereas it provided little additional cosmological information relative to the substantially higher computational cost.

For comparison, the other statistic $\overline{S}_1(j \in [6,7])$ retains moderate cosmological information but is more strongly affected by tracer bias ($\chi^2_{\nu, \rm bias} = 9.8$). The normalized power spectrum $P_{\rm norm}(k)$, while effectively suppressing bias ($\chi^2_{\nu, \rm bias} \sim 0.1$), captures almost no cosmological signal ($\chi^2_{\nu, \rm cos} \sim 0.2$), limiting its usefulness for the parameter sensitivity.

In summary, only $R^{\rm wst}(j \in [3,7])$ simultaneously satisfies $\chi^2_{\nu, \rm cos} \approx 6$ and $\chi^2_{\nu, \rm bias} \sim 1$--a regime not achieved by any other method examined. These results highlight $R^{\rm wst}$ as a promising tool for future galaxy survey analyses, offering robust cosmological sensitivity with effective bias mitigation.

Further research is needed to model tracer bias with more realistic galaxy models, such as HOD and SHAM, and to incorporate assembly bias. Additionally, several systematics in the scattering transform warrant deeper investigation, including the impact of survey geometry, redshift errors in CSST-like surveys, and the combined analysis of $R^{\rm wst}$ with other statistical measures, to enhance our understanding of LSS, refine cosmological insights, and improve the discriminative power of statistical methods in cosmology.

\begin{acknowledgments}
This work was supported by the National SKA Program of China (2020SKA0110401, 2020SKA0110402, 2020SKA0110100), the National Key R\&D Program of China (2020YFC2201600), the National Natural Science Foundation of China (12373005, 12473097, 12073088), the China Manned Space Project (CMS-CSST-2021: A02, A03, B01), the Fundamental Research Funds for the Central Universities, Sun Yat-sen University (No. 24qnpy122), and the Guangdong Basic and Applied Basic Research Foundation (2024A1515012309). We thank Prof. Y.P. Jing for providing the \texttt{CosmicGrowth} cosmological simulation suite. We also acknowledge the Beijing Super Cloud Center (BSCC) and Beijing Beilong Super Cloud Computing Co., Ltd. (http://www.blsc.cn/) for providing HPC resources that substantially supported this study.
\end{acknowledgments}

\appendix
\section{Linear bias cancellation in WST $m$-mode ratios}\label{app:bias}

In this Appendix, we demonstrate that the WST $m$-mode ratios cancel linear bias terms. 
For a 3D overdensity field with a linear tracer bias, 
\begin{equation}
I_0(\bm{x}) = b_1 \, \delta(\bm{x}),
\end{equation}
From Eq.~\ref{eq:S_1}, the first-order WST coefficients scale as
\begin{equation}
S_1(j, l, m) = \big\langle | I_0 \otimes \Psi_{j, l}^{m} |^q \big\rangle \propto b_1^q\,.
\end{equation}

The first-order $m$-mode ratios (defined in Eq.~\ref{eq:R1}) is insensitive to linear bias because
\begin{equation}
U_1 = | I_0 \otimes \Psi_{j,l}^{m} | \propto b_1.
\end{equation}
Substituting this scaling into the definition of $R^{\rm wst}(j, l)$, the factor $b_1^q$ cancels in the ratio, rendering the first-order $m$-mode ratio independent of linear bias.

Similarly, the second-order WST coefficients are given by
\begin{equation}
S_2(j', j, l, m) = \left\langle \big| U_1 \otimes \Psi_{j',l}^{m} \big|^q \right\rangle,
\end{equation}
so that $S_2 \propto b_1^q$. Constructing the second-order m-mode ratios (see Eq.~\ref{eq:R2}) again cancels the linear bias factor exactly, demonstrating that both first- and second-order $m$-mode ratios are robust to linear tracer bias.

\section{Analytical illustration of $m$-mode ratios for a directionally modulated field}\label{app:dir}

To illustrate the behavior of the WST $m$-mode ratios, we consider, without loss of generality, a 3D cosmological overdensity field with a directional modulation along a given axis. The field can be written as
\begin{equation}
I(\bm{x}) = A \left[ 1 + \epsilon\, Y_l^m(\hat{\bm{x}}) \right], \quad 0 < \epsilon \ll 1,
\end{equation}
where $Y_l^m$ are the spherical harmonics, $\hat{\bm{x}} = \bm{x}/|\bm{x}|$, $A$ is the mean amplitude, and $\epsilon$ controls the modulation strength.

From the first-order WST coefficients in Eq.~\ref{eq:S_1} (with $q=1$), substituting the modulated field $I(\bm{x}) = A[1 + \epsilon\, Y_l^m(\hat{\bm{x}})]$ gives
\begin{equation}
S_1(j,l,m) = \Big\langle \Big|\, A \underbrace{\int \Psi_{j,l}^{m}(\bm{y})\, d^3y}_{C_0} 
+ A \epsilon \underbrace{\int Y_l^m(\hat{\bm{y}})\, \Psi_{j,l}^{m}(\bm{x}-\bm{y})\, d^3y}_{R(j,l,m,\bm{x})} \,\Big| \Big\rangle,
\end{equation}
where the first term corresponds to the response of the wavelet to a uniform field, and the second term represents the convolution of the directional modulation with the wavelet. Taking the spatial average $\langle \cdot \rangle$ and expanding to first order in $\epsilon \ll 1$, we define
$R(j,l,m) \equiv \big\langle R(j,l,m,\bm{x}) \big\rangle$,
which captures the contribution from the directional modulation while $C_0$ is independent of both $\epsilon$ and the specific spherical harmonic $Y_l^m$.

The first-order $m$-mode ratios are then
\begin{equation}
R^{\rm wst}(j,l) = \frac{1}{l} \sum_{m=0}^{l-1} \frac{S_1(j,l,m+1)}{S_1(j,l,m)} 
\simeq 1 + \frac{\epsilon}{l\, C_0} \sum_{m=0}^{l-1} \big[ R(j,l,m+1) - R(j,l,m) \big],
\end{equation}
where the expansion is to first order in $\epsilon \ll 1$. Here, $C_0$ represents the response of the wavelet $\Psi_{j,l}^m$ at scale $j$ and angular degree $l$ to a uniform field, and is independent of both the modulation amplitude $\epsilon$ and the specific spherical harmonic $Y_l^m$. The terms $R(j,l,m)$ encode the contributions from the directional modulation.

Similarly, for the second-order WST coefficients defined in Eq.~\ref{eq:S_1}, we have
\begin{equation}
U_1(\bm{x}) = \big| I_0 \otimes \Psi_{j,l}^{m} \big| \simeq C_0 + \epsilon\, R(j,l,m,\bm{x}),
\end{equation}
where $R(j,l,m,\bm{x})$ encodes the contribution from the directional modulation before spatial averaging. Substituting this into the second-order coefficients,
\begin{equation}
S_2(j',j,l,m) = \Big\langle \big| U_1 \otimes \Psi_{j',l}^{m} \big| \Big\rangle
\simeq \Big\langle \big| C_0 + \epsilon \, R(j,l,m,\bm{x}) \otimes \Psi_{j',l}^{m} \big| \Big\rangle.
\end{equation}

After taking the spatial average and expanding to first order in $\epsilon \ll 1$, we define
\begin{equation}
R'(j',j,l,m) \equiv \Big\langle R(j,l,m,\bm{x}) \otimes \Psi_{j',l}^{m} \Big\rangle,
\end{equation}
which represents the contribution of the directional modulation at scale $j'$ after convolving with the second wavelet.

Then the second-order $m$-mode ratio becomes
\begin{equation}
R_2^{\rm wst}(j',j,l) = \frac{1}{l} \sum_{m=0}^{l-1} \frac{S_2(j',j,l,m+1)}{S_2(j',j,l,m)} 
\simeq 1 + \frac{\epsilon}{l\, C_0} \sum_{m=0}^{l-1} \Big[ R'(j',j,l,m+1) - R'(j',j,l,m) \Big].
\end{equation}

In summary, to first order in $\epsilon$, both the first- and second-order $m$-mode ratios, $R^{\rm wst}$ and $R_2^{\rm wst}$, are sensitive only to directional modulations in the field, with the isotropic component $C_0$ largely canceling in the ratios. When $\epsilon = 0$ (no directional modulation), both ratios reduce exactly to unity. For $\epsilon \neq 0$, they deviate from unity proportionally to the modulation amplitude, providing a direct measure of the field’s anisotropy. While $R^{\rm wst}$ captures the immediate angular modulation of the field, $R_2^{\rm wst}$ additionally encodes scale-to-scale correlations between anisotropic features.

\section{Impacts of second-order WST and multipole selection}
\label{app:s2}
According to the results in Sect.~\ref{sec:R}, where $j \in [3,7]$ was identified as the optimal range balancing high sensitivity to cosmological parameters and low sensitivity to tracer bias, we adopt $(j=3, j'=4)$ to construct the second-order statistic $R_2^{\rm wst}$. We further evaluate $R_2^{\rm wst}$ for multipole combinations ranging from $l \in [1,3]$ to $l \in [1,2]$. This comparison serves as a useful indicator of the performance of second-order statistics relative to their first-order counterparts under different multipole selections.

Tab.~\ref{tab:chi2_R^2} summarizes the reduced and total chi-square values for cosmological sensitivity and tracer bias under high-density apodization. For $l \in [1,3]$, $R_2^{\rm wst}(j=3,j'=4)$ yields $\chi^2_{\nu,\mathrm{cos}}\approx 4.5$, which is substantially lower than $R^{\rm wst}(j=3)$ ($\sim 25.2$) but only slightly higher than $R^{\rm wst}(j=4)$ ($\sim 3.7$), while $\chi^2_{\nu,\rm bias}$ remains close to unity and within $1\sigma$. Moreover, for $l \in [1,2]$, $R_2^{\rm wst}$ produces $\chi^2_{\rm cos}\approx 12.3$, still much lower than $R^{\rm wst}(j=3)$, with negligible bias impact. These results indicate that incorporating $R_2^{\rm wst}$ does not provide a significant improvement in cosmological parameter sensitivity over the main results presented in the text. 

On the other hand, when comparing the $R^{\rm wst}$ results for $l \in [1,2]$ and $l \in [1,3]$, we find that $\chi^2_{\nu,\mathrm{cos}}$ increases from $16.25$ to $25.15$ for $j=3$ and from $2.95$ to $3.67$ for $j=4$, while the bias sensitivity remains minor ($1.07$ for $j=3$ and $0.47$ for $j=4$). These results indicate that including the additional $l=3$ mode can improve cosmological parameter sensitivity without significantly affecting bias. However, considering the increased computational cost associated with higher-order multipoles, we restrict our analysis to $l=1,2$ for a conservative evaluation. The inclusion of higher-order multipoles will be addressed in future study.

\begin{table}
\centering 
\caption{Results for the reduced chi-square ($\chi^2_\nu$) and total chi-square ($\chi^2$), quantifying the sensitivity to cosmological parameters and tracer bias under the high-density apodization. Comparisons are shown for combining three multipoles ($l \in [1,3]$) versus two multipoles ($l \in [1,2]$), as well as for $R^{\rm wst}_2$ with fixed indices $(j=3, j'=4)$ versus $R^{\rm wst}$ with either $j=3$ or $j=4$. The $1\sigma$ uncertainty, $\sigma_\nu$, derived from the degrees of freedom of $\chi^2_\nu$, is reported in the last row. }\label{tab:chi2_R^2} 
\begin{tabular}{l|ccc|ccc}
\hline
\hline
\multirow{2}{*}{\textbf{Parameter}} & \multicolumn{3}{c|}{$l\in [1,3]$} & \multicolumn{3}{c}{$l\in[1,2]$} \\
\cline{2-7}

& $R_2^{\rm{wst}}(j=3,j'=4)$ & $R^{\rm{wst}}(j=3)$ & $R^{\rm{wst}}(j=4)$ & $R_2^{\rm{wst}}(j=3,j'=4)$ & $R^{\rm{wst}}(j=3)$ & $R^{\rm{wst}}(j=4)$ \\
\hline
\multirow{1}{*}{$\chi^2_{\nu,\mathrm{cos}}$/$\chi^2_{\mathrm{cos}}$} 
      & 4.47/13.41 & 25.15/75.44 & 3.67/11.02 & 6.13/12.26 & 16.25/32.5 & 2.95/5.91\\ 
\hline
\multirow{1}{*}{$\chi^2_{\nu,\mathrm{bias}}$/$\chi^2_{\mathrm{bias}}$} 
      & 0.24/0.72  &  1.07/3.2 & 0.47/1.41 & 0.14/0.28 & 0.23/0.47 &  0.26/0.52\\
\hline
\multirow{1}{*}{$\sigma_{\nu}$} 
      & 0.82 & 0.82 & 0.82 & 1 & 1 & 1 \\
\hline
\hline
\end{tabular}
\end{table}

\nocite{*}
\bibliography{sample7}{}
\bibliographystyle{aasjournalv7}

\end{document}